\documentclass[pre,aps,twocolumn,showpacs,floatfix,10pt,
superscriptaddress]{revtex4}
\usepackage{graphicx}
\usepackage{amsmath, amsthm, amssymb}
\newcommand{\be}{\begin{equation}}
\newcommand{\ee}{\end{equation}}
\newcommand{\bea}{\begin{eqnarray}}
\newcommand{\eea}{\end{eqnarray}}

\begin{document}

\title{Reversible limit of processes of heat transfer}
\author{J\"{u}rgen F. Stilck}
\affiliation{Instituto de F\'{i}sica and National Institute of Science
and
Technology for Complex Systems, Universidade Federal Fluminense, Av.
Litor\^anea s/n, 24210-346 Niter\'oi, RJ, Brazil}
\email{jstilck@if.uff.br}
\author{Rafael Mynssem Brum}
\affiliation{Instituto de F\'{i}sica,
Universidade Federal Fluminense, Av.
Litor\^anea s/n, 24210-346 Niter\'oi, RJ, Brazil}

\date{\today}

\begin{abstract}
We study a process of heat transfer between a body of heat capacity $C(T)$ 
and a sequence 
of $N$ heat reservoirs, with temperatures equally spaced between an initial 
temperature $T_0$ and a final temperature $T_N$. The body and the heat 
reservoirs 
are isolated from the rest of the universe, and the body is brought in thermal
contact successively with reservoirs of increasing temperature. 
We determine the  
change of entropy of the composite thermodynamic system in the total process
in which the temperature of the body changes from $T_0$ to $T_N$.
We find that for large values of $N$ the total change of entropy
of the composite process is proportional to $(T_N-T_0)/N$, but eventually
a non-monotonic behavior is found at small values of $N$.
\end{abstract}


\maketitle

\section{\label{sec:intro}Introduction}
One of the original formulations of the second law of thermodynamics, 
due to Clausius,
states that ``no process is possible whose only result is the transfer of heat 
from a body of lower temperature to a body of higher temperature''. 
As discussed by
Fermi in his book \cite{fermi}, since as this statement was made the concept of 
temperature was empirical, to find out which of two bodies has the 
higher temperature
one had to put them in thermal contact and find out in which sense 
the heat flows, it
will flow spontaneously from the body of higher temperature to the other. So, 
Clausius's statement could be rephrased into ``If heat flows spontaneously from
body A to body B, no process is possible whose only result is the transfer of
heat from body B to body A''. To be more precise, let us isolate the two bodies 
from the rest of the universe, to assure that they will not exchange 
heat or other
forms of energy with the environment, this is implicit in the term 
`only result'. We may notice in this statement that already the 
idea of irreversibility is present: if heat flows spontaneously 
from A to B, it will 
not flow spontaneously from B to A. In other words, the process 
of transfer of heat 
between two bodies which are isolated from the environment is 
{\em irreversible}.

The second law off thermodynamics says that not all processes which conserve 
energy, as required by the first law, occur in nature. In many cases 
(if not all),
if a given process in an isolated system occurs, the inverse process 
does not. This
idea was further formalized by the introduction of the concept of 
{\em entropy}, so
that processes in isolated systems which lead to a decrease of the entropy are 
not allowed. In particular, in the process of heat transfer mentioned above, the
entropy of the composite isolated system increases if the heat flows from the
body of higher temperature to the one of lower temperature, but would 
{\em decrease} 
in the inverse process, so that only the further is allowed. 
It is interesting to 
recall that in his paper of 1885, Clausius summarized the two first laws of
thermodynamics in just two sentences: ``The energy of the 
universe is constant. The
entropy of the universe tends to a maximum'' \cite{clausius}. 
We may notice that 
the universe is the only composite thermodynamic system which is isolated by 
definition.

The entropy in equilibrium thermodynamics is a state function, so that it is 
defined for equilibrium states of the system. When two bodies
of different temperatures are brought into thermal contact, the composite system
in general will go through a sequence of non-equilibrium states, but if we wait
long enough, it will finally reach a new equilibrium state, where 
both bodies of the composite have the
same temperature, and in this state the entropy is larger than 
it was in the initial
state. If we now imagine a process which happens at a very slow rate in time
when compared to the relaxation time of the system \cite{callen}, we may reach
a situation close to the one in which all intermediate states of the system are
equilibrium states. In this limit the process is called quasi-static, 
and therefore
the whole process may be represented as a path in the space of variables which 
define the equilibrium states of the system. In the particular case in which the
entropy of the final state of the quasi-static process is equal to the 
entropy of the initial state, and therefore to the entropies of 
all intermediate 
states, this process will be {\em reversible}. 

The process we will study in some detail here is the transfer of 
heat in isolated systems composed by a heat reservoir (constant temperature)
and a body whose heat capacity is described by a function $C(T)$. 
The initial temperature 
of the body is $T_0$ and its final temperature is $T_N$. The
change of the
temperature of the body is accomplished by putting it
in thermal contact
with a sequence of $N$ heat reservoirs whose temperatures are equally 
spaced between 
$T_0$ and $T_N$. For a finite number of reservoirs, each of the sub-processes 
consists of heat transfer between two bodies at different temperatures and 
is therefore irreversible, leading to an increase of the entropy of the
composite isolated system. In the limit $N \to \infty$, however, 
each sub-process will
contribute with a negligible increase of entropy and, what may not be obvious
initially, the total variation of entropy vanishes as well, so that this is a
concrete example of a limit which leads to a reversible, quasi-static process
of heat transfer.

This problem, of considerable pedagogical interest, has been studied before in 
the literature. For a constant heat capacity, such as is found for ideal 
gases, Calkin and Kiang \cite{ck83} have 
shown for a cyclic process ($T_0 \to T_N \to T_0$), that the total change of
entropy vanishes as $1/N$ in the limit $N \to \infty$ because, although the
number of sub-processes increases, the increase of entropy in each of them 
is proportional to  $1/N^2$ in this limit. A similar situation is proposed in 
the exercise 4.4-6 in Callen's book \cite{callen}. The same process was studied 
also by Thomsen
and Bers \cite{tb96}, again for the case of constant heat capacity and 
discussing in detail 
that it is always possible, given any positive value $\epsilon$, 
to choose a value
of $N$ for which the increase of entropy is smaller than $\epsilon$. 
This argument
is then repeated for the inverse process, thus characterizing the process in
the limit $N \to \infty$ as reversible.

Here we generalize the problem in two ways: the 
heat capacity of the system is no longer constant and we develop the increase
of entropy as a series in $1/N$, so that we find the corrections to the leading 
term. We also show that some symmetries in the result which are valid in the
asymptotic limit close to reversibility are not longer present for finite values
of $N$. In particular, we show that, although in the large $N$ limit the 
increase of entropy decreases monotonically with $N$, this may not happen at
lower numbers of heat reservoirs. Also, the increase of entropy for the
direct process ($T_0 \to T_N$) may be different of the same quantity for the 
inverse process ($T_N \to T_0$). The determination of the change of entropy is
given in section \ref{sec:entropy} and applications to particular cases 
and final comments may be found in section \ref{sec:discussion}.

\section{\label{sec:entropy}Determination of the change of entropy}
\begin{figure}[h]
\includegraphics[scale=0.6]{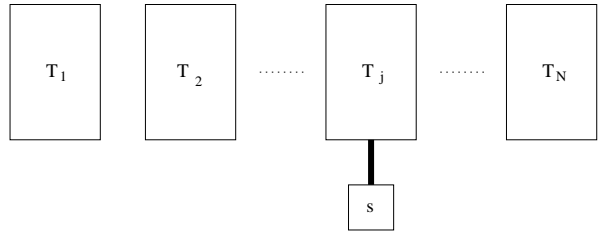}
\caption{Illustration of the $j$'th process of heat transfer, where 
the system $s$ is
in thermal contact with the heat reservoir at temperature $T_j$, so that its 
temperature is changed from $T_{j-1}$ to $T_j$. The pair 
system-reservoir at $T_j$ is
isolated, as are all the other reservoirs.} 
\label{process}
\end{figure}

Let us define the problem in more detail. Our composite system 
consists of the body, 
whose heat capacity is $C(T)$ and which will undergo a temperature change 
$\Delta T=T_N-T_0$, and a set of $N$ heat reservoirs, at equally 
spaced temperatures, 
so that the temperature of the reservoir $j$ ($j=1,2,3,\ldots,N$) is equal to
$T_j=T_0+j\,\Delta T/N$. The body is brought into thermal contact with the heat 
reservoirs in order of increasing values of $j$ until its temperature equals
the one of the reservoir and therefore is increased bi $\Delta T/N$. In the 
first sub-process, for example, the body starts with temperature 
$T_0$ and ends with
temperature $T_0+\Delta T/N$. After the $N$'th process, the body 
will reach the final
temperature $T_N$. This sequence of heat transfer processes is 
illustrated in figure \ref{process}. Actually, one could argue that the time
required to reach equilibrium for each sub-process of heat transfer would
be infinite, and for this reason, Thomsen and Bers, in their discussion of 
a similar process \cite{tb96}, have chosen the temperature of the $j$'th
reservoir to be $T_0+(j+1)\Delta T/N$, thus assuring that the equilibration 
time of the system with each reservoir is finite. Here, for simplicity, we 
will not use this improved version of the process.

If a the body, at temperature $T$, receives a quantity of heat 
$dQ_b$, its temperature 
will change by $dT=dQ_b/C(T)$. The change of entropy of the body in 
the whole sequence of heat transfer processes is:
\be
\Delta S_b=\int_{T_0}^{T_N} \frac{dQ_b}{T}= \nonumber \\
\int_{T_0}^{T_N} \frac{C(T)}{T}\,dT.
\ee
The change of entropy of all reservoirs in the sequence of processes will be:
\bea
\Delta S_r(N)&=&\sum_{j=1}^N \Delta S_r(N,j)=\sum_{j=1}^N \frac{1}{T_j}
\int_{T_{j-1}}^{T_j}dQ_r=\nonumber \\
&&-\sum_{j=1}^N \frac{1}{T_j} \int_{T_{j-1}}^{T_j} C(T)\,dT,
\eea
where in the last passage, we recall that in each sub-process, the body and the 
reservoir are isolated, so that the heat received by the reservoir is given
by $dQ_r=-dQ_b=-C(T)\,dT$. We thus may write the total change of entropy as:
\bea
\Delta S(N)&=&\Delta S_b(N)+\Delta S_r(N)= \nonumber \\
&&\int_{T_0}^{T_N} \frac{C(T)}{T}\,dT-\sum_{j=1}^N \frac{1}{T_j}\int_{T_{j-1}}^{T_j}
C(T)\,dT.
\label{deltas}
\eea
If it is possible to perform both integrations, this expression above
becomes:
\be
\Delta S(N)=S_b(T_N)-S_b(T_0)-\sum_{j=1}^N \frac{1}{T_j}[U(T_{j})-
U(T_{j-1})],
\label{deltas1}
\ee
where $S_b(T)=\int C(T)/T\,dT$ and $U(T)=\int C(T)\,dT$.
It is convenient to rewrite the expression for the change of entropy as:
\be
\Delta S(N)=\sum_{j=1}^N\int_{T_{j-1}}^{T_j} \kappa_j(T,N)\phi(T)\,dT,
\label{deltas2}
\ee
where $\kappa_j(T,N)=1-T/T_j$ and $\phi(T)=C(T)/T$. It is now easy to see that,
since stability implies that $\phi(T) \ge 0$, we must have $\Delta S \ge 0$,
as required by the second law of thermodynamics. Let us show this in 
some detail.
Suppose that $T_N>T_0$, so that the temperature of the body increases in the
process. It is then clear that $\kappa_j(T,N) \ge 0$ and as a consequence the
entropy of the composite system increases. The same happens in the inverse
process, where the temperature of the body decreases from $T_N$ to $T_0$. 
In this case the change if entropy is:
\be 
\Delta S^\prime(N)=-\sum_{j=1}^N\int_{T_{j-1}}^{T_j} \kappa_{j-1}(T,N)\phi(T)\,dT,
\ee
and since now $\kappa_{j-1}(T,N) \le 0$ in the range of integration, again the
entropy increases. It may be worth noticing that the difference of the entropy
increases between the processes with raising and lowering of the temperature
of the body $\delta S(N)=\Delta S^\prime(N)-\Delta S(N)$ is:
\be
\delta S(N)=\sum_{j=1}^N\int_{T_{j-1}}^{T_j} \left[\left(
\frac{1}{T_{j-1}}+\frac{1}{T_j}\right)T-2\right]\phi(T)\,dT,
\label{deinv}
\ee
and the sign of this expression is not defined in general. 

\begin{figure}
\includegraphics[scale=0.3]{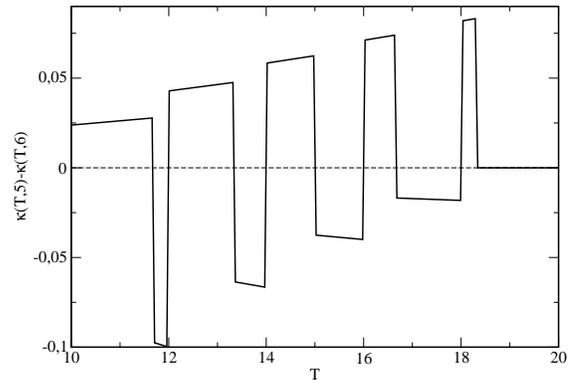}
\caption{The function $\kappa(T,5)-\kappa(T,6)$ for initial 
temperature $T_0=10$ and final temperature $T_N=20$.} 
\label{deltakappa}
\end{figure}

Another general property of the entropy increase is its dependence 
of the number
of reservoirs $N$. As we will see below, it vanishes in the 
limit $N \to \infty$ 
of a quasi-static process. The question we will address is if the change of
the entropy decreases monotonically with $N$. For convenience, we will define
the function:
\be
\kappa(T,N)=\sum_{j=1}^N \Theta(T-T_{j-1})\Theta(T_j-T)\,\kappa_j(T,N),
\ee
where $\Theta(T)$ is the step function. The function $\kappa(T,N)$ 
is defined in 
the whole temperature range $[T_0,T_N]$ and has a sawtooth pattern, 
with $N$ maxima 
of decreasing values. The expression for the entropy increase 
Eq. (\ref{deltas2})
may then be cast into the following form:
\be
\Delta S(N)=\int_{T_0}^{T_N} \kappa(T,N)\phi(T)\,dT.
\ee
The difference
of entropy increases for two values of $N$, ($N_2>N_1$) is:
\be
\Delta S(N_1)-\Delta S(N_2)=\int_{T_0}^{T_N}[\kappa(T,N_1)-
\kappa(T,N_2)]\phi(T)\,dT.
\label{deltas12}
\ee
It is straightforward to see that if $N_2$ is a multiple of $N_1$ the 
difference of $\kappa$ functions is non-negative, and therefore 
$\Delta S(N_1)-\Delta S(nN_1) 
\ge 0$ for any integer $n$. For example, we have $\Delta S(1)>\Delta S(2)$.
This, however, is no longer true 
in general if the ratio
$N_2/N_1$ is not an integer. As an example, in figure \ref{deltakappa} we show 
the difference of $\kappa$ functions for $N_1=5$ and $N_2=6$, and we
notice that this difference assumes negative values in part of the temperature
domain. We thus reach the 
conclusion that, depending on the function $\phi(T)$, the entropy 
increase for $N_2$
heat reservoirs may be larger than the one for $N_1$ reservoirs, if 
$1<N_2/N_1<2$.

We now proceed obtaining the general asymptotic behavior of $\Delta S(N)$
for $N \gg 1$. Expanding the last integral in the expression for the 
increase of entropy, Eq. (\ref{deltas}), which corresponds to
the amount of heat $Q_j$ exchanged by the body with reservoir $j$, in a Taylor
series around the temperature $T_j$, we get:
\be
Q_j=-\sum_{i=1}^\infty \frac{(-1)^i}{i\!}\left(\frac{\Delta T}{N}\right)^i
C^{(i-1)}(T_j),
\ee
where a superscript between parenthesis means a derivative of the corresponding
function of the order of the superscript. We may now write the total change 
of entropy as:
\bea
\Delta S(N)&=&\int_{T_0}^{T_N} \frac{C(T)}{T}\,dT+ \nonumber \\
&&\sum_{i=1}^\infty 
\frac{(-1)^i}{i\!}\left(\frac{\Delta T}{N}\right)^i\sum_{j=1}^N
\frac{C^{(i-1)}(T_j)}{T_j}.
\label{deltase}
\eea

We now proceed transforming the sum over the reservoirs into an integral, 
using Euler-MacLaurin's expansion \cite{knopp}:
\bea
\sum_{i=m}^n f(i)&=&\int_m^nf(x)\,dx+\frac{1}{2}[f(n)+f(m)]+ \nonumber \\
&&\sum_{k=1}^{k_{max}} \frac{B_{2k}}{(2k)!}\left[f^{(2k-1)}(n)-f^{(2k-1)}(m)\right]+
\nonumber \\
&&R_{k_{max}},
\label{eml}
\eea
where $B_{2k}$ are the Bernoulli numbers ($B_2=1/6$, $B_4=-1/30$, $B_6=1/42$,
$B_8=-1/30$, $\ldots$. See \cite{bn} for a list of the numbers). The series in
the right hand side is divergent in many cases, since the Bernoulli numbers
increase very fast for larger values of $k$. There are procedures to 
evaluate the 
rest $R_{k_{max}}$ \cite {knopp} and we will discuss this issue in the appendix.
The sum we will convert into an integral is:
\be
G_i=\sum_{j=1}^N \frac{C^{(i-1)}(T_j)}{T_j}=\sum_{j=0}^N \frac{C^{(i-1)}(T_j)}{T_j}-
\frac{C^{(i-1)}(T_0)}{T_0},
\ee
and applying the expansion (\ref{eml}) we find:
\bea
G_i&=&\frac{N}{\Delta T}\int_{T_0}^{T_N} \frac{C(T)^{(i-1)}}{T}\,dT+
\nonumber \\
&&\frac{1}{2}
\left[\frac{C^{(i-1)}(T_N)}{T_N}-\frac{C^{(i-1)}(T_0)}{T_0}\right]+ \nonumber \\
&&\sum_{k=1}^{k_{max}} \frac{B_{2k}}{(2k)!}\left(\frac{\Delta T}{N}\right)^{2k-1}
\times \nonumber \\
&&\left[\left(\frac{C^{(i-1)}(T_N)}{T_N}\right)^{(2k-1)}- \right. \nonumber \\
&&\left.\left(\frac{C^{(i-1)}(T_0)}{T_0}\right)^{(2k-1)}\right]+ 
R_{k_{max}}.
\label{gi}
\eea

Now we may substitute this last expression into Eq. (\ref{deltase}) and collect
the terms in the change of entropy of the body in the sequence of processes
in powers of $1/N$. We notice, as expected, than the term of 
order $0$ cancels, so that, in the
limit $N \to \infty$ the whole process becomes reversible. In the appendix, 
we proceed calculating all coefficients $\sigma_i$ of the expansion
\be
\Delta S(N)=\sum_{i=1}^{i_{max}} \sigma_i \left(\frac{1}{N}\right)^i,
\label{entropexp}
\ee
but here we 
we will limit ourselves to the leading term, which is:
\be
\sigma_1=\frac{T_N-T_0}{2}\left[\int_{T_0}^{T_N} \frac{C^{(1)}(T)}{T}\,dT-
\left(\frac{C(T_N)}{T_N}+\frac{C(T_0)}{T_0} \right)\right].
\label{lt}
\ee
Integrating by parts, we finally obtain the leading term of the expansion:
\be
\Delta S(N) \approx \frac{T_N-T_0}{2N}\int_{T_0}^{T_N}
\frac{C(T)}{T^2}\,dT,
\ee
valid in the limit $N \gg 1$.

Since $C(T) \ge 0$, we notice that the leading term vanishes only in
the trivial case where $C(T)=0$ in the whole range of temperatures and
there is no heat transfer, so that the asymptotic behavior as $1/N$ in the
increase of entropy, as was found in the particular case of an ideal
gas \cite{ck83}, is universal. We also remark that in the asymptotic
regime the entropy increase is invariant if the temperatures are 
interchanged and also monotonically decreasing with $N$, properties
which are not true in general for small values of the number of 
reservoirs.

\section{\label{sec:discussion}Discussions and conclusion}
\begin{figure}
\includegraphics[scale=0.3]{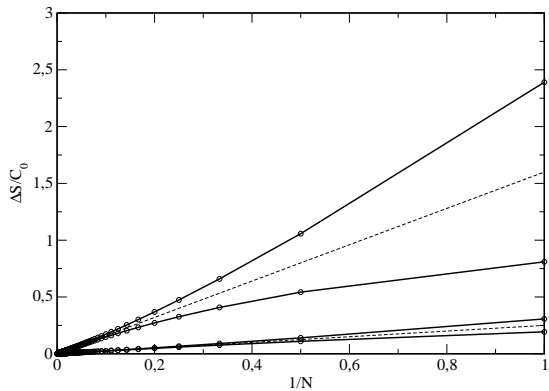}
\caption{Increase of entropy for a system with constant heat capacity $C_0$ 
(ideal gas) as a function
of the inverse number of reservoirs $1/N$. The four curves, in the upward
order, correspond to ratios of temperatures $f=T_N/T_0=2,\;1/2,\;5$ and
$1/5$. The dashed straight lines are the asymptotic behavior for small 
$1/N$. 
}
\label{sgi}
\end{figure}

We will now apply the results of the preceding section to some 
particular cases of physical interest. We start with the ideal gas,
which has a constant heat capacity $C(T)=C_0$. Applying Eq. (\ref{deltas})
to this particular case, the increase of the entropy will be:
\be
\frac{\Delta S(N)}{C_0}=\ln f-\frac{f-1}{N}\sum_{j=1}^N
\frac{1}{1+\frac{j}{N}(f-1)},
\ee
where $f=T_N/T_0$. The leading term in this case is:
\be
\frac{\Delta S(N)}{C_0} \approx \frac{1}{2N}\left(\sqrt{f}-
\sqrt{1/f}\right)^2.
\ee
In figure \ref{sgi} we show some curves of the entropy increase as a function
of $1/N$, for different values of the ratio of temperatures $f$, as well as 
the dashed straight lines which represent the asymptotic behavior. Notice that,
as expected, the same asymptotic behavior is found for $f=2$ and $f=1/2$.
If we consider two processes, one with ratio $f_1>1$ and the other with
$f_2=1/f_1<1$, that is, related by interchange of the temperatures, both will
have the same asymptotic behavior, as may be seen in the particular cases
depicted in the figure. Also, the increase of entropy in the second 
process, where the
body is cooled, is always larger than the one in the first process. Another
way to state the same property is that, as can be seen in the curves,
the increase of entropy is a concave function of 
$1/N$ for processes where the body is heated and it is a convex
function when the temperature of the body decreases. In the asymptotic 
regime these properties may be obtained the next term of the 
Euler-MacLaurin expansion, as will be discussed in the appendix. Also,
the increase of entropy for this particular case is a monotonically 
decreasing function of $N$. 

In most cases the variation of entropy decreases monotonically
with $N$, a property which seems also intuitive, since by decreasing
the temperature steps in a certain sense the composite process gets 
closer to a reversible process. For the ideal gas this is always
the case. We have seen above, however, that this may not be
generally true. If we look at the Eq. (\ref{deltas12}) for the 
difference of entropy increases for different values of $N$, a
non-monotonic result is possible if the function $\phi(T)$ has one 
or more peaks at the temperatures where $\kappa(T,N_1)-
\kappa(T,N_2)$ is negative. One system which could possibly
show a non-monotonic
behavior is a two-state system \cite{callen}, since its heat capacity
displays a peak, sometimes called ``Schottky hump''. However, we found
that this is not the case, the reason is that the peak is too wide to
produce a non-monotonic behavior.

We turn our attention to a situation where the width of the peak in the
heat capacity may be changed by adjusting a parameter. The simple expression
we will use for the heat capacity is:
\be
C(T)=C[\tau(2-\tau)]^m,
\label{c-peak}
\ee 
where $\tau=T/T_c$, so that $T_c$ is the temperature where the maximum
of the heat capacity is located, and $C$ is a constant with the 
dimension of entropy. Although this heat capacity does not
correspond to a particular physical system, it is convenient for 
analytic calculations and consistent with the third law of 
thermodynamics. Of course, at least for odd values of the 
parameter $m$, it is valid only if $\tau$ is in the range $[0,2]$, so 
we will restrict ourselves to this range of temperatures. As the parameter 
$m$ grows, the peak in $C(T)$ narrows, and our main interest in this 
system is that, for sufficiently large values of $m$, non-monotonic
behavior is found in $\Delta S(N)$. 

\begin{figure}
\includegraphics[scale=0.3]{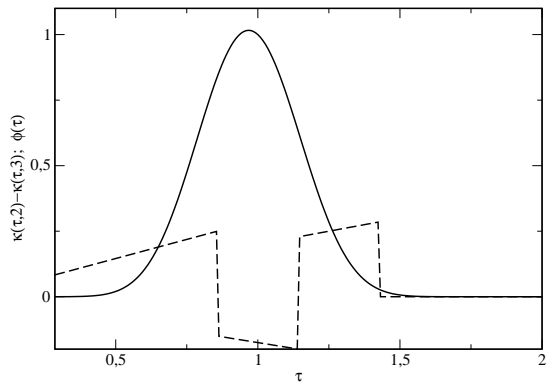}
\caption{$\kappa(\tau,2)-\kappa(\tau,3)$ (dashed line) and the function
$\phi(\tau)=C(\tau)/(C\tau)=[\tau(2-\tau)]^{16}/\tau$. The initial 
temperature is $\tau_0=0.286$ and the final temperature is $\tau_N=2.0$
}
\label{kappa-c}
\end{figure}

In figure \ref{kappa-c} we show, in the same graph, the
function $\kappa(\tau,2)-\kappa(\tau,3)$ and the function 
$\phi(\tau)=C(\tau)/(C\tau)$,
given by Eq. (\ref{c-peak}) with $m=16$. As is apparent in the graph, with
the choices of initial ($\tau_0=0.286$) and final ($\tau_N=2.0$) temperatures
we made, the maximum of the peak in $C(T)$ is close to the center of the 
negative peak in $\kappa(\tau,2)-\kappa(\tau,3)$, so that the difference
$\Delta S(2)-\Delta S(3)$ is minimized, since this difference is given
by Eq. (\ref{deltas12}). 
In table \ref{peak-deltas23} we list the results 
for $[\Delta S(3)-\Delta S(2)]/C$ 
for different values of $m$, with the choices above for $\tau_0$ and $\tau_N$,
and it is clear that a non-monotonic behavior is found for $m \ge 16$. Finally,
we found some evidence that a local maximum in the entropy could also be found
at $N=4$, but if this actually happens for this particular form
of heat capacity, it will be at rather large values of $m$, which lead
to numerical problems. We remark that, in order to reduce numerical errors, 
in all examples we discuss the integrations involved in the calculation of
the entropy increase Eq. (\ref{deltas}) are performed analytically, as is
done in Eq. (\ref{deltas1}).

\begin{table}
\begin{ruledtabular}
\begin{tabular}{cc}
$m$&$[\Delta S(3)-\Delta S(2)]/C$\\
\hline
$14$ & $-3.158309 \times 10^{-3}$ \\
$15$ & $-3.427008 \times 10^{-4}$ \\
$16$ &  $2.185200 \times 10^{-3}$ \\
$17$ &  $4.463576 \times 10^{-3}$ \\
\end{tabular}
\end{ruledtabular}
\caption{$[\Delta S(3)-\Delta S(2)]/C$ for different values of the parameter
$m$ for a system with heat capacity given by Eq. (\ref{c-peak}). The 
temperature range of the process is $\tau_0=0.286$, $\tau_N=2.0$.
}
\label{peak-deltas23}
\end{table}

As discussed before, we expect, qualitatively,  the non-monotonic 
behavior to be enhanced if the peak in the  heat capacity is more 
pronounced. This leads us, in this final example, to a system 
which undergoes a continuous phase 
transition, since in this case a singularity at the critical temperature 
$T_c$ is found in the heat capacity, of the form 
$C(T) \approx |1-T/T_c|^{-\alpha}$, described by the critical exponent 
$\alpha$ \cite{yeomans}. To the singular behavior of the heat capacity, 
usually regular contributions should be added, so that we will use the
simple expression:
\be
C(T)=C \tau |1-\tau|^{-\alpha},
\label{c-phaset}
\ee
where we define the reduced dimensionless temperature $\tau=T/T_c$ and 
which is compatible with the third law of thermodynamics. It should be
mentioned that since the exchange of heat when the temperature
of the body crosses the critical temperature has to be finite, 
we should restrict the critical exponent $\alpha$ to values smaller than $1$. 
It is easy to perform the necessary integrations in this case, so that 
the entropy increase will be given by Eq. (\ref{deltas1}) with:
\be
\frac{S_b(\tau)}{C}=\left\{ \begin{array}{ll}
-\frac{(1-\tau)^{1-\alpha}}{1-\alpha}, & \mbox{$\tau<1$,} \\
  & \\
\frac{(\tau-1)^{1-\alpha}}{1-\alpha} & \mbox{$\tau \ge 1$,}
\end{array}
\right.
\ee
and
\be
\frac{U(\tau)}{CT_c}=\left\{ \begin{array}{ll}
-\frac{(1-\tau)^{1-\alpha}}{(1-\alpha)(2-\alpha)}[1+(1-\alpha)\tau], &
\mbox{$\tau<1$} \\
 & \\
\frac{(\tau-1)^{1-\alpha}}{(1-\alpha)(2-\alpha)}[1+(1-\alpha)\tau], &
\mbox{$\tau \ge 1.$}
\end{array}
\right.
\ee

\begin{figure}
\includegraphics[scale=0.3]{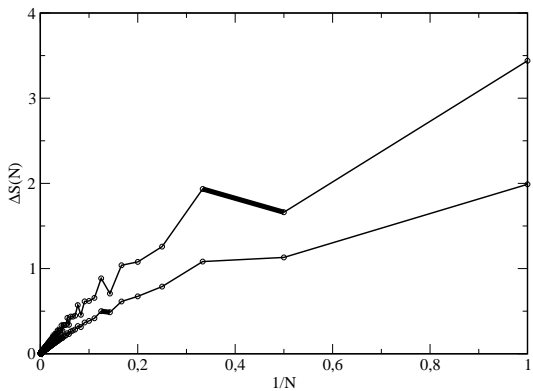}
\caption{Values of $\Delta S(N)$ as function of $1/N$ for a body with
a continuous phase transition (Eq. (\ref{c-phaset})). The upper curve is
for $\alpha=0.7$ and the lower one for $\alpha=0.5$. The pairs of data 
at the end of the thick lines are the first ones, in the order of increasing
values of $N$, where a increase of $\Delta S(N)$ is found. These results
are for $\tau_0=0.2$ and $\tau_N=2.2$.
}
\label{deltasn-phaset}
\end{figure}

We would expect that the non-monotonic behavior of the entropy should be
enhanced as the value of the critical exponent $\alpha$ increases. This
is actually true, but at variance with what is seen in the preceding example
of a peak in the heat capacity, the local maxima of the entropy increase,
for lower values of $\alpha$, appear at larger values of $N$. In figure 
\ref{deltasn-phaset} we see results of the increase of entropy for
two values of the exponent, $\alpha=0.7$ and $\alpha=0.5$. It is apparent
that, for $\alpha=0.7$ we already have an increase of the change of 
entropy between $N_1=2$
and $N_2=3$, the segment with a thicker line in the curve. Many other pairs
of successive results with the same property are found for larger values 
of $N$. For $\alpha=0.5$, the first pair where the change of entropy increases
is seen at $N_1=7$, $N_2=8$.

\begin{figure}
\includegraphics[scale=0.3]{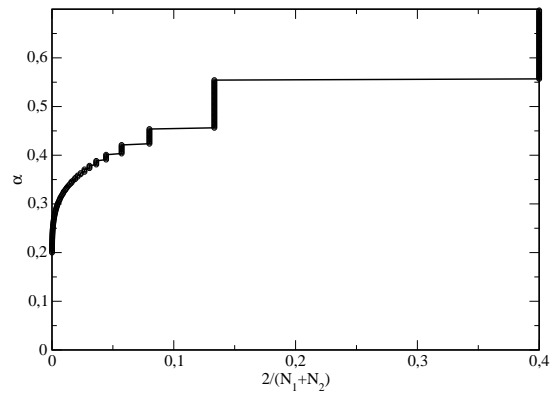}
\caption{Values of the critical exponent $\alpha$ for which the first
increase of $\Delta S(N)$ if found between $N_1$ and $N_2=N_1+1$, as
function of $2/(N_1+N_2)$. The data are for $\tau_0=0.2$ and $\tau_N=2.2$.
}
\label{nonm}
\end{figure}

For a given a value of the critical exponent 
$\alpha$, we may find $N_1$, the lowest value of $N$ such that 
$\Delta S(N_2)>\Delta S(N_1)$, where $N_2=N_1+1$. In figure \ref{nonm}, the
exponent $\alpha$ is plotted as a function of $2/(N_1+N_2)$, and we
notice that, at least for large values of $\alpha$, that the function shows
steps. The results shown are calculated for $\tau_0=0.2$ and $\tau_N=2.2$. 
Thus, for example, if $\alpha>0.555$, we notice that the first increase
of $\Delta S(N)$ is seen between $N_1=2$ and $N_2=3$, while if $\alpha$ 
is in the range $[0.555,0.454]$ this increase is seen at $N_1=7$, $N_2=8$,
if $\alpha$ is further decreased the increase shifts to $N_1=12$, $N_2=13$,
and so  on. This pattern, although qualitatively remaining the same, 
changes quantitatively for a different temperature range. 
We notice an interesting behavior for small values of $\alpha$, which
leads to the question if, for any positive $\alpha$ an increase if 
$\Delta S(N)$ is found for a finite value of $N$. Due to round-off errors,
we could not address this question numerically. However, since the 
coefficients $\sigma_i$ in the Euler-MacLaurin expansion
Eq. (\ref{entropexp}) diverge in this case for $i>1$ and any positive
$\alpha$, the answer to this question should be positive. We discuss this 
point somewhat more in the appendix.

\begin{figure}
\includegraphics[scale=0.3]{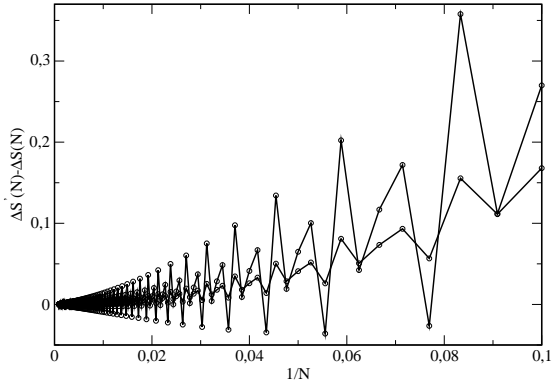}
\caption{Difference in the increases of entropy in the cooling and the
heating processes $\Delta S^\prime(N)-\Delta S(N)$, for $\tau_0=0.2$ and
$\tau_N=2.2$. The curve with larger oscillations is for $\alpha=0.7$ and
the other one for $\alpha=0.5$.
}
\label{dsspplot}
\end{figure}

Finally, we notice that, in opposition to what was found in the first two
examples we studied, when the body undergoes a continuous phase transition
it is possible that $\Delta S(N)>\Delta S^\prime (N)$, that is, the increase 
of entropy in the heating process is larger than the one in the cooling
process, a possibility which was anticipated in the discussion after Eq.
(\ref{deinv}). In figure \ref{dsspplot} we present results for the 
difference in entropy increases $\Delta S^\prime(N)-\Delta S(N)$, for 
$\alpha=0.7$ and $\alpha=0.5$, as functions of $1/N$. The temperature 
interval is $\tau_0=0.2$, $\tau_N=2.2$ Although in most cases 
this difference is positive, for $\alpha=0.7$ already for $n=13$ we get
a negative result, and a set of values of $N$ above this one have the 
same property. For $\alpha=0.5$ the effect is smaller, and the first
occurrence of a negative result is for $N=88$. Again it would be interesting 
to study more details of these results at large values of $N$, but
round-off errors prevent this to be done numerically.

A final remark on this example is that, if $\tau \ll1$, 
$C(\tau) \approx C\tau$, so that it will show a behavior similar to a gas 
of non-interacting fermions at low temperature. In this case, if the initial 
temperature $\tau_0$ approaches $0$, the coefficient $\sigma_1$ diverges
as a logarithm. In other words, in a heating process starting at $\tau_0=0$, 
the tangent to the curve $\Delta S \times 1/N$ at the origin is vertical.
This curve is concave, since $\sigma_2$ is negative. The even more unphysical 
cooling process with vanishing final temperature shows a very uncommon 
behavior: $\Delta S(N)$ diverges for any finite $N$ and vanishes as 
$N \to \infty$. 

In conclusion, we notice that in general the process of heat transfer
between a body and a sequence of heat reservoirs, is a very rich
physical situation for understanding of the second law of thermodynamics 
and the concept of a reversible process. In particular, if the body 
undergoes a continuous phase transition in the composite process, some
curious phenomena appear , which at first seem to defy common sense, 
but of course none of them violates thermodynamics.

\appendix*
\section{Euler-MacLaurin expansion}
\label{app}
Here we will develop the terms of the Euler-MacLaurin expansion
in some detail. 
If we substitute Eq. (\ref{gi}) into Eq. (\ref{deltase}) and collect
the terms in the change of entropy of the body in the sequence of processes
in powers of $\Delta T/N$. We notice, as expected, than the term of 
order $0$ cancels, so that, in the
limit $N \to \infty$ the whole process becomes reversible. The coefficients 
we obtain for the change of entropy Eq. (\ref{entropexp}) are:
\bea
&&\sigma_i=(-1)^i(T_N-T_0)^i \left\{\frac{-1}{(i+1)\!}
\int_{T_0}^{T_N}\frac{C^{(i)}(T)}{T}
\,dT+ \right. \nonumber \\
&& \left.
+\frac{1}{2i\!}\left(\frac{C^{(i-1)}(T_N)}{T_N}-\frac{C^{(i-1)}(T_0)}{T_0}\right)+
\right.\nonumber \\
&&\left.
\sum_{k=1}^{[i/2]}\frac{(-1)^{2k-1}B_{2k}}{(i+1-2k)!(2k)!} \times \right.
\nonumber \\
&&\left.\left[\left(\frac{C^{(i-2k)}(T_N)}{T_N}\right)^{(2k-1)}-\right.\right. 
\nonumber \\
&&\left.\left.\left(\frac{C^{(i-2k)}(T_0)}{T_0}\right)^{(2k-1)}\right]\right\},
\; i=1,2,3,\ldots
\label{coefs}
\eea
The two first coefficients of this series, after integrating by parts,
are:, are:
\be
\sigma_1=\frac{T_N-T_0}{2}\int_{T_0}^{T_N} \frac{C(T)}{T^2}\,dT,
\ee
and
\bea
\sigma_2&=&-\frac{(T_N-T_0)^2}{12}\left[4\int_{T_0}^{T_N}
\frac{C(T)}{T^3}\,dT+
\right.\nonumber \\
&&\left.\frac{C(T_N)}{T_N^2}-\frac{C(T_0)}{T_0^2}\right].
\eea
We notice that $\sigma_1 \ge 0$, and vanishes only if the 
temperature interval is zero or if $C(T)=0$ in the whole interval. In 
other words, if there is a non-zero heat exchange between the body and
the reservoirs, $\sigma_1$ is positive.  If this
coefficient could be negative, the second law would be violated. It is
also invariant with respect to a permutation of the temperatures, as was
already noted in the examples discussed above. The sign of the second
coefficient $\sigma_2$ is not fixed, and, as happens with all coefficients of
even order, it switches under permutation of the temperatures. The odd 
order coefficients do not change their sign under this permutation. 

In the particular case of an ideal gas, where $C(T)=C_0$, the coefficients 
are:
\bea
\sigma_1&=&-C_0\,\frac{T_N-T_0}{2}\left(\frac{1}{T_N}-\frac{1}{T_0}\right), \\
\sigma_{2k}&=&C_0\,\frac{B_{2k}(T_N-T_0)^{2k}}{2k}\left(\frac{1}{T_N^{2k}}-
\frac{1}{T_0^{2k}}\right), 
\eea
where the last expression is valid for k=1,2,3,\ldots.
We notice that, with the exception of the dominant term, only 
terms with even powers of $1/N$ appear in the expansion for this 
particular case. For the heating process $T_N>T_0$, $\sigma_2<0$, so that 
the $\Delta S$ is a concave function of $1/N$ in the limit $N \to \infty$.
For the cooling process $T_N<T_0$, the function is convex. These properties
can be verified in the numerical results presented in figure \ref{sgi}. 

\begin{figure}[t!]
\includegraphics[scale=0.3]{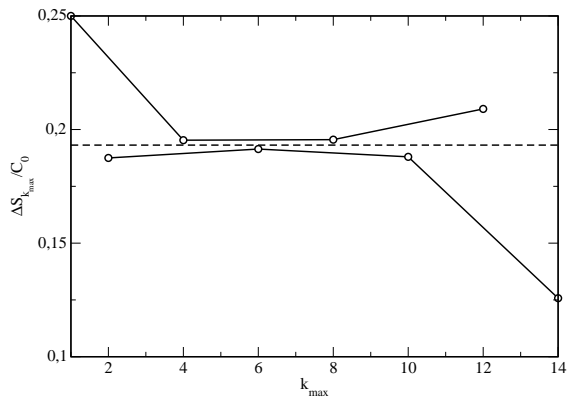}
\caption{Increase of entropy for a system with constant heat capacity $C_0$ 
as a function
the order of truncation of the asymptotic expansion $i_{max}$. The ratio of 
temperatures $T_N/T_0$ is equal to 2 and the number of reservoirs is $N=1$. 
The exact value corresponds to the dashed line.} 
\label{deltasimax}
\end{figure}

Very often the expansion in the Euler-MacLaurin summation 
(\ref{eml}) does not converge. We find that even for $N=1$ 
the truncated asymptotic
values are close to the exact ones at rather low orders. However, truncating
at higher orders does not necessarily lead to results closer to the exact ones.
For example, for $N=1$ the result closest to the exact one is found truncating
the series at $k_{max}=6$. We notice that higher orders of truncation lead
to wrong results, as may be seen in figure \ref{deltasimax}, where the increase
of entropy calculated using the expansion Eq. (\ref{entropexp}), for $N=1$ and 
$T_N/T_0=2$, is shown for several orders of truncation. The results oscillate
around the exact value at successive orders. The best result is obtained for 
$i_{max}=6$ and if we truncate the expansion at larger orders we observe that
they start deviating from the exact value.

In the table \ref{diffent}, we present more results of such calculations. 
We notice that quite 
accurate estimates of the entropy change may be obtained if the series 
is truncated
at the most favorable order. The question of determining the order 
which minimizes
the rest $R$ in the Euler-MacLaurin expansion Eq. (\ref{eml}) is 
rather technical
and we refer to the specialized literature for details, but
in general it may be assured that, if $f(x)$ in Eq. (\ref{eml}) has always
the same sign and if the function and all its derivatives tend monotonically
to $0$ as $x \to \infty$, which are valid for $f(N)=\Delta S(N)$, then
the rest $R_{k_{max}}$ is of the same order and has the same sign of the first
neglected term \cite{knopp}.

\begin{table}[t!]
\begin{tabular}{ccccc}
\hline
\hline
$T_N/T_0$ &  $N$ & $k_{max}$ & $|\Delta S-\Delta S_{k_{max}}|/\Delta S$ & 
$\Delta S$ \\
\hline
 $1/3$ & $1$ & $2$ & $6.8311 \times 10^{-2}$ & $0.9014$ \\
 $1/3$ & $2$ & $6$ & $4.3650 \times 10^{-3}$ & $0.4014$ \\
 $1/3$ & $3$ & $8$ & $2.5634 \times 10^{-4}$ & $0.2538$ \\
 $1/3$ & $4$ &$12$ & $1.2824 \times 10^{-5}$ & $0.1847$ \\
 $1/3$ & $5$ &$16$ & $6.4910 \times 10^{-7}$ & $0.1450$ \\
 $1/3$ & $6$ &$18$ & $3.0760 \times 10^{-8}$ & $0.1192$ \\
 $1/2$ & $1$ & $6$ & $5.6735 \times 10^{-3}$ & $0.3069$ \\
 $1/2$ & $2$ &$12$ & $1.6897 \times 10^{-5}$ & $0.1402$ \\
 $1/2$ & $3$ &$18$ & $4.0671 \times 10^{-8}$ & $0.0765$ \\
   $2$ & $1$ & $6$ & $9.0135 \times 10^{-3}$ & $0.1931$ \\
   $2$ & $2$ &$12$ & $2.1570 \times 10^{-5}$ & $0.1098$ \\
   $2$ & $3$ &$18$ & $4.7960 \times 10^{-8}$ & $0.0765$ \\
   $3$ & $1$ & $2$ & $1.4255 \times 10^{-1}$ & $0.4319$ \\
   $3$ & $2$ & $6$ & $6.6046 \times 10^{-3}$ & $0.2653$ \\
   $3$ & $3$ &$10$ & $3.4783 \times 10^{-4}$ & $0.1907$ \\
   $3$ & $4$ &$12$ & $1.5940 \times 10^{-5}$ & $0.1486$ \\
   $3$ & $5$ &$16$ & $7.7345 \times 10^{-7}$ & $0.1217$ \\
   $3$ & $6$ &$18$ & $3.5619 \times 10^{-8}$ & $0.1030$ \\
\hline
\hline
\end{tabular}
\caption{Body with constant heat capacity ($C(T)=C_0$). Values of the 
relative difference between the increase of entropy calculated
directly ($\Delta S$) and the one which follows truncating the series Eq. 
(\ref{entropexp}) ($\Delta S_{k_{max}}$). Several values for $T_N/T_0$ and 
$N$ are considered, 
and the optimal order of truncation $i_{max}$ is given.}
\label{diffent}
\end{table}

Finally, it is clear from Eq. (\ref{coefs}) for the expansion coefficients,
that for the case of a singular heat capacity at a critical temperature $T_c$
($C(T)=C\tau(1-\tau)^\alpha$, with $\tau=T/T_c$), all but the first coefficients
diverge if the temperature interval includes the critical temperature. This 
fact may account for the rather peculiar behavior found numerically in the 
large $N$ limit.

\end{document}